\documentclass[a4paper,11pt]{article}

\usepackage{jinstpub} 
\usepackage{multirow}
\usepackage{import}
\usepackage{subcaption}
\usepackage{xcolor}

\title{\boldmath
Studies in Pulse Shape Discrimination for an Optimized ASIC Design}

\author[a,1]{B. Boxer \note{Corresponding author.}}
\author[a]{B. Godfrey}
\author[b]{C. Grace}
\author[a]{J. Johnson}
\author[a]{R. Khandwala}
\author[a]{M. Tripathi}

\affiliation[a]{University of California, Davis, Department of Physics,\\ Davis, CA 95616-5270, USA}
\affiliation[b]{Lawrence Berkeley National Laboratory (LBNL),\\ Berkeley, CA 94720-8099, USA}

\emailAdd{bboxer@ucdavis.edu}

\abstract{The continued advancements of Silicon Photomultipliers (SiPMs) have made them viable photosensors for low recoil energy Pulse Shape Discrimination (PSD) between fast neutron and gamma interactions when coupled to an appropriate scintillator. At the same time, the large number of channels in a typical array calls for the development of low-cost and low-power electronics. A custom integrated circuit (ASIC) is an ideal solution for this purpose. To assess the requirements for such an ASIC, studies were performed using two scintillators, Stilbene and EJ-276, coupled to a 6 x 6 mm SiPM from Onsemi. We demonstrate that both scintillators are viable for performing PSD for interaction energies from 100 keV to several MeV while optimizing the integration periods used in the PSD  metric. These measurements inform the design parameters of the ASIC under development.}


\begin{document}
\maketitle
\flushbottom

\section{Introduction}
\label{sec:intro} 
Plastic and organic crystalline scintillators have demonstrated the capability of performing Pulse Shape Discrimination (PSD) to distinguish between different types of ionizing radiation \cite{Brooks_plastic}. This capability is attributed to the dependency of these scintillators' stopping power on the type of incident ionizing radiation. This has led to their widespread use in devices in which they are coupled to an optical sensor, such as a PhotoMultiplier Tube (PMT) or Silicon PhotoMuliplier (SiPM). The primary applications of such devices are neutron physics, nonproliferation efforts, and nuclear security.
There is a demonstrable need for compact electronics that operate as front-end systems for high-density array readout for specialized end-use applications, such as fast neutron/gamma PSD. Additionally, various scintillators are being developed and improved for such PSD. Existing readout systems rely on fast digitization and data processing to perform PSD. This method is both expensive and dissipates large amounts of power. Hence, there is a need for low-power front-end systems capable of providing the required functionality. Application-specific integrated circuits (ASICs) are naturally low-power devices and have the flexibility for implementing custom circuits such as real-time analog PSD methods. Further, when produced in large quantities, ASICs are cost-effective compared to commercial off-the-shelf digitizers and processors. 
Low-powered devices are highly desirable as they widen their applicability to mobile devices and make them viable for deployment in cryogenic environments. Additionally, the use of an ASIC can potentially reduce the number of components in a multi-channel system. This is an important consideration when developing and constructing compact multichannel devices such as a compact neutron camera.

This article presents PSD studies of two scintillators, stilbene and EJ-276, coupled to an Onsemi J-series SiPM array to inform the design of an ASIC developed for onboard real-time analog PSD. This ASIC incorporates a scalable multi-channel low-power readout system with a low-noise front-end, real-time PSD, and a highly tunable synthesized digital core.
Additionally, the ASIC has an adjustable transimpedance gain and two different techniques for the programmable digital delay lines. The ASIC is designed to provide an input dynamic range of order 1 MeV energy depositions, depending on the light yield of specific scintillators. The high level of programmability on-chip is maintained in the synthesized digital core. Current targeted end-use applications include neutron cameras and active neutron-tagging systems for nuclear recoil calibration for dark matter and neutrino experiments. This work will not discuss in detail the design parameters of the ASIC. The effort here measures and optimizes parameters specific to a scintillator type, such as the partial and total integration periods required to perform a `\textit{Q-ratio}' approach for PSD, as described below. These measurements were used to inform the programmable features of this highly flexible ASIC.

Solution-grown trans-stilbene has become a common choice for scintillator-based PSD due to its high light output, 10700 $\pm$ 1700 photons per MeV \cite{SB_KIM2013133}. However, stilbene has several drawbacks; it is both fragile and susceptible to environmental damage; it is also relatively expensive and poses difficulties when producing large crystals without defects. Plastic scintillators, such as EJ-276 developed by Eljen Technology, are easily machinable, far less susceptible to environmental damage, inexpensive when produced in large quantities, and can be produced without significant defects. EJ-276 has a notably lower light yield than stilbene, 9800 $\pm$ 1000 photons per MeV \cite{EJ276_Grodzicka_2020}, but has significant differences in the intensities of the fast component of the emission for fast neutrons compared to gammas. The relative intensities and decay constants for each component of both scintillators' emissions are given in Table \ref{tab:emission-components}, which indicates that for both scintillators, more photons are expected in the tail of a waveform for a fast neutron than a gamma. Therefore, a Q-ratio assessment is a suitable approach for performing PSD of these particle types. Typically, a Q-ratio assessment is performed as a peak-to-tail comparison, whereas this work uses the ratio between a partial and total integration of the waveform \cite{QRATIO_SELLIN_2003}. The start of the integration is defined as $t_0$, the partial integration ends at $t_1$ and the total at $t_2$:

\begin{equation} \label{eq:PSD}
\mathrm{PSD}=\frac{\int_{t_0}^{t_1} Waveform(t)\; \it{dt}}{\int_{t_0}^{t_2} Waveform(t)\;\textit{dt}}.
\end{equation}

Two metrics can be applied to assess the viability of a given combination of partial and total integration periods. The first metric defines a Figure Of Merit (FOM) for the distribution of the PSD values obtained for neutrons and gammas that takes into account the degree of separation between the two distributions in terms of their means, $\mu_{\rm{n/\gamma}}$ and standard deviations, $\sigma_{\rm{n/\gamma}}$:
\begin{equation} \label{eq:FOM}
    \mathrm{FOM}=\frac{\mu_{\gamma}-\mu_{n}}{2.355\left(\sigma_{\gamma}+\sigma_{n}\right)}.
\end{equation}
Two Gaussian distributions can be considered as sufficiently separated when the FOM is greater than 1.27; this being the value equivalent to 3($\sigma_\gamma$ + $\sigma_n$)/2.355($\sigma_\gamma$ + $\sigma_n$), where 2.355$\sigma_{n/\gamma}$ is the full width at half maximum of the corresponding peak \cite{FOM}.
The second metric assesses the probability of misidentifying a gamma event as a neutron (leakage) while maintaining a 99.9\% acceptance for neutron identification. Applications for PSD towards the identification of neutrons are typically conducted in scenarios with a gamma rate that is orders of magnitude greater than the neutron rate. This is an important metric to consider as it directly indicates the false positive rate of identifying gammas as neutrons.
It is calculated by determining the percent of gamma events with a PSD value at or below the PSD corresponding to the 99.9\% bound of the neutron distribution. Therefore, the optimal combination of partial and total integration periods would maximize the FOM while minimizing the gamma leakage. Applying these two metrics to various combinations of partial and total integration periods makes it possible to determine the optimal combination.

\begin{table}[tb]
\caption{Parameters used to fit the emission components of the scintillation for EJ-276 \cite{EJ276_Grodzicka_2020} and stilbene \cite{SB_KIM2013133}.} 
\label{tab:emission-components}
\resizebox{\textwidth}{!}{%
    \begin{tabular}{cccccccccc}
    \hline
    \hline
    \multirow{2}{*}{Scintillator} & \multirow{2}{*}{Particle} & \multicolumn{2}{c}{Fast} & \multicolumn{2}{c}{Intermediate} & \multicolumn{4}{c}{Delayed} \\ \cline{3-10} 
     &  & \multicolumn{1}{c}{\begin{tabular}[c]{@{}c@{}}Decay const.\\ {[}ns{]}\end{tabular}} & \begin{tabular}[c]{@{}c@{}}Intensity\\ \%\end{tabular} & \multicolumn{1}{c}{\begin{tabular}[c]{@{}c@{}}Decay const. \\ {[}ns{]}\end{tabular}} & \begin{tabular}[c]{@{}c@{}}Intensity\\ \%\end{tabular} & \multicolumn{1}{c}{\begin{tabular}[c]{@{}c@{}}Decay const.\\ {[}ns{]}\end{tabular}} & \multicolumn{1}{c}{\begin{tabular}[c]{@{}c@{}}Intensity\\ \%\end{tabular}} & \multicolumn{1}{c}{\begin{tabular}[c]{@{}c@{}}Decay const.\\ {[}ns{]}\end{tabular}} & \begin{tabular}[c]{@{}c@{}}Intensity\\ \%\end{tabular} \\ \hline
    \multirow{2}{*}{EJ-276} & Gamma & \multicolumn{1}{c}{4.0} & 71 & \multicolumn{1}{c}{16} & 12 & \multicolumn{1}{c}{98} & \multicolumn{1}{c}{8} & \multicolumn{1}{c}{690} & 9 \\ 
     & Fast neutron. & \multicolumn{1}{c}{3.9} & 47 & \multicolumn{1}{c}{18} & 13 & \multicolumn{1}{c}{106} & \multicolumn{1}{c}{13} & \multicolumn{1}{c}{800} & 27 \\ \hline
    \multirow{2}{*}{Stilbene} & Gamma & \multicolumn{1}{c}{5.3} & 95 & \multicolumn{1}{c}{21} & 3 & \multicolumn{1}{c}{135} & \multicolumn{1}{c}{2} & \multicolumn{1}{c}{N/A} & N/A \\  
     & Fast neutron & \multicolumn{1}{c}{5.0} & 95 & \multicolumn{1}{c}{28} & 4 & \multicolumn{1}{c}{253} & \multicolumn{1}{c}{1} & \multicolumn{1}{c}{N/A} & N/A \\ \hline\hline
    \end{tabular}%
}
\end{table}

\section{Method and Energy Calibration}
\label{sec:method}  
In order to assess the PSD capabilities of both EJ-276 and stilbene, a simple test bed was constructed: a dark box, a 60035 Onsemi J-series SiPM (36.8 mm$^2$ active area) connected to a 50 $\Omega$ terminated breakout board, CAEN DT5725 digitizer, and a computer interface. Onsemi SiPMs have an interesting design feature. Along with the standard output (SOUT), there is also a capacitively coupled ultra-fast output (FOUT), which is derived from the fast internal switching of the microcells induced by the turn-on of the avalanche. This pulse has a full-width half maximum (FWHM) of 3 ns for a single photon, giving a far faster impulse than the SOUT, which has a 50 ns shaping time. The breakout board was used to power the SiPM array and read out the SOUT and FOUT waveform.
\begin{figure}[b]
    \centering
    \includegraphics[width=0.98\textwidth]{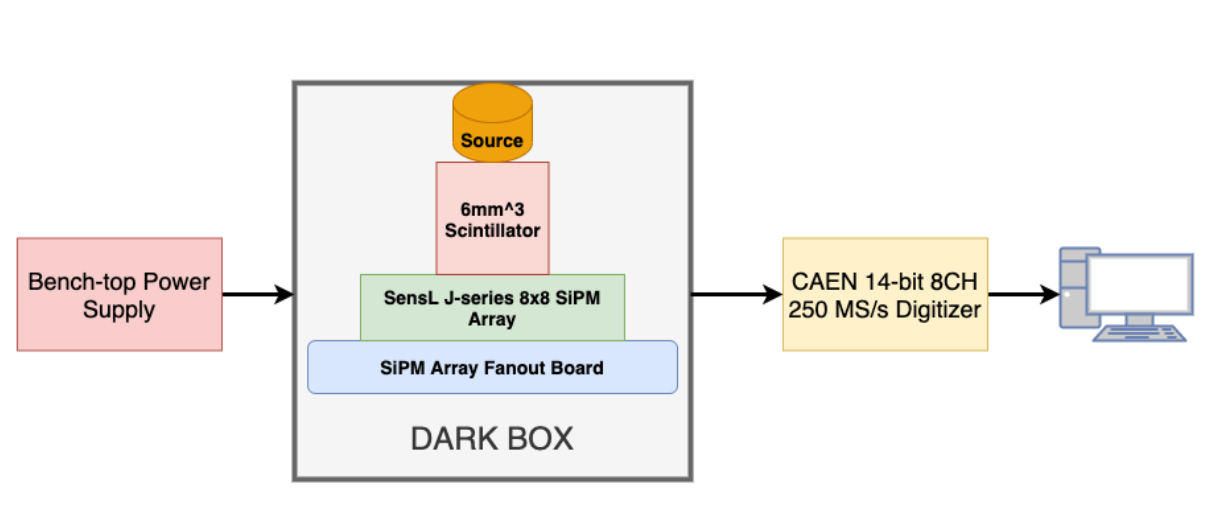}
    \caption{Diagram of the desk top test bed}
    \label{fig:my_label}
\end{figure}

These waveforms were digitized at 250 MS/s with a 14-bit digital resolution. A 2 $\rm{\mu}$s acquisition window with a 15\% pre-trigger window was captured. Due to the low amplitude of the FOUT signal, acquisitions were triggered using SOUT. Triggering was set sufficiently high enough to ensure the rate of dark counts was $<$ 50 mHz. The SiPM was biased at 29 V for these tests (4.7 V overvoltage). Both scintillators were 6mm cubes, ensuring the scintillator covered the entire active area of the active SiPM. Both scintillators were wrapped in PTFE reflective coatings. The scintillators were procured from Inrad Optics \cite{SB_DS} and Eljen Technology \cite{EJ276_DS}, respectively.

\begin{figure*}[htb!]
    \centering
    \includegraphics[width=0.45\textwidth]{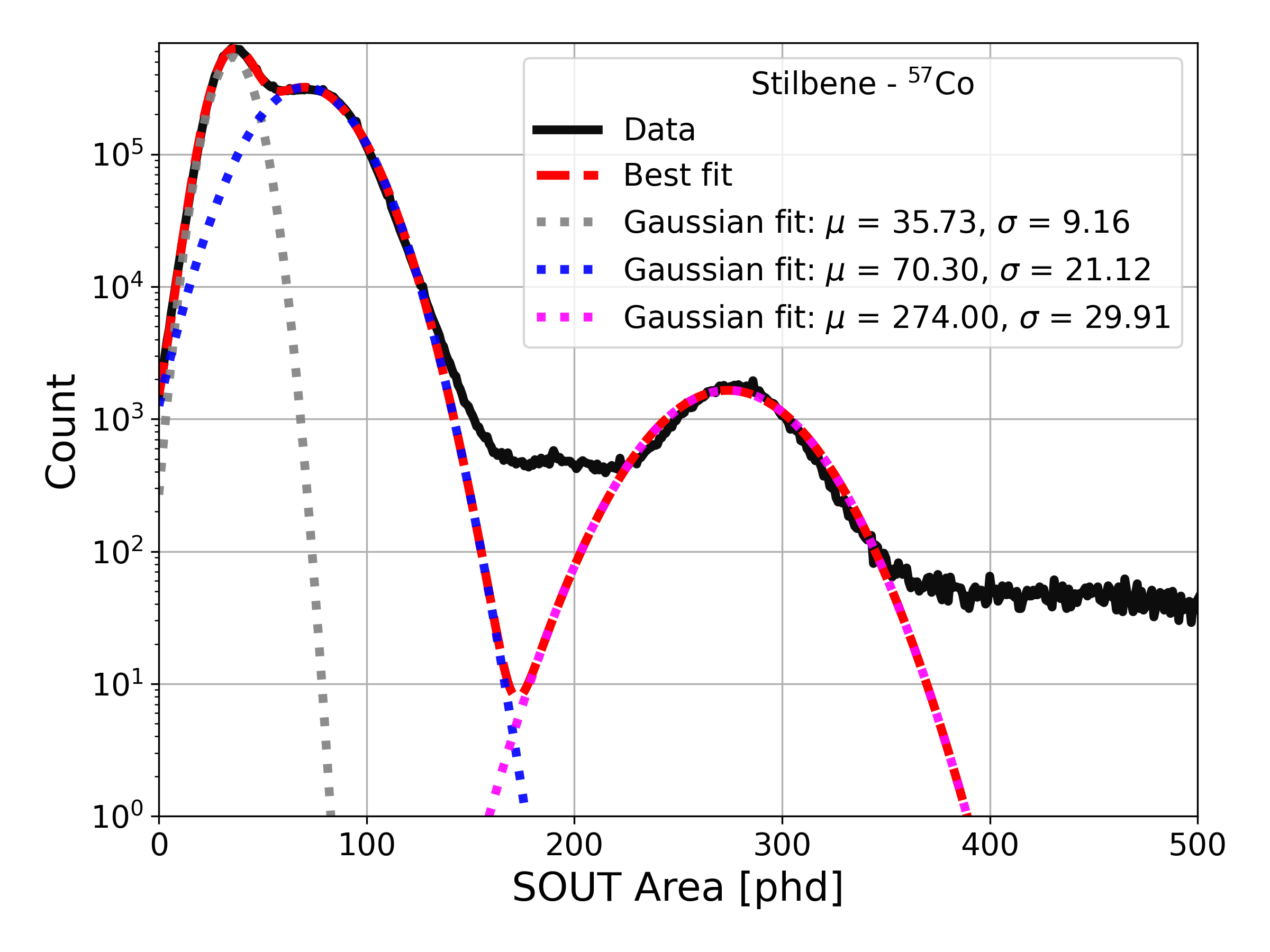}
    \includegraphics[width=0.45\textwidth]{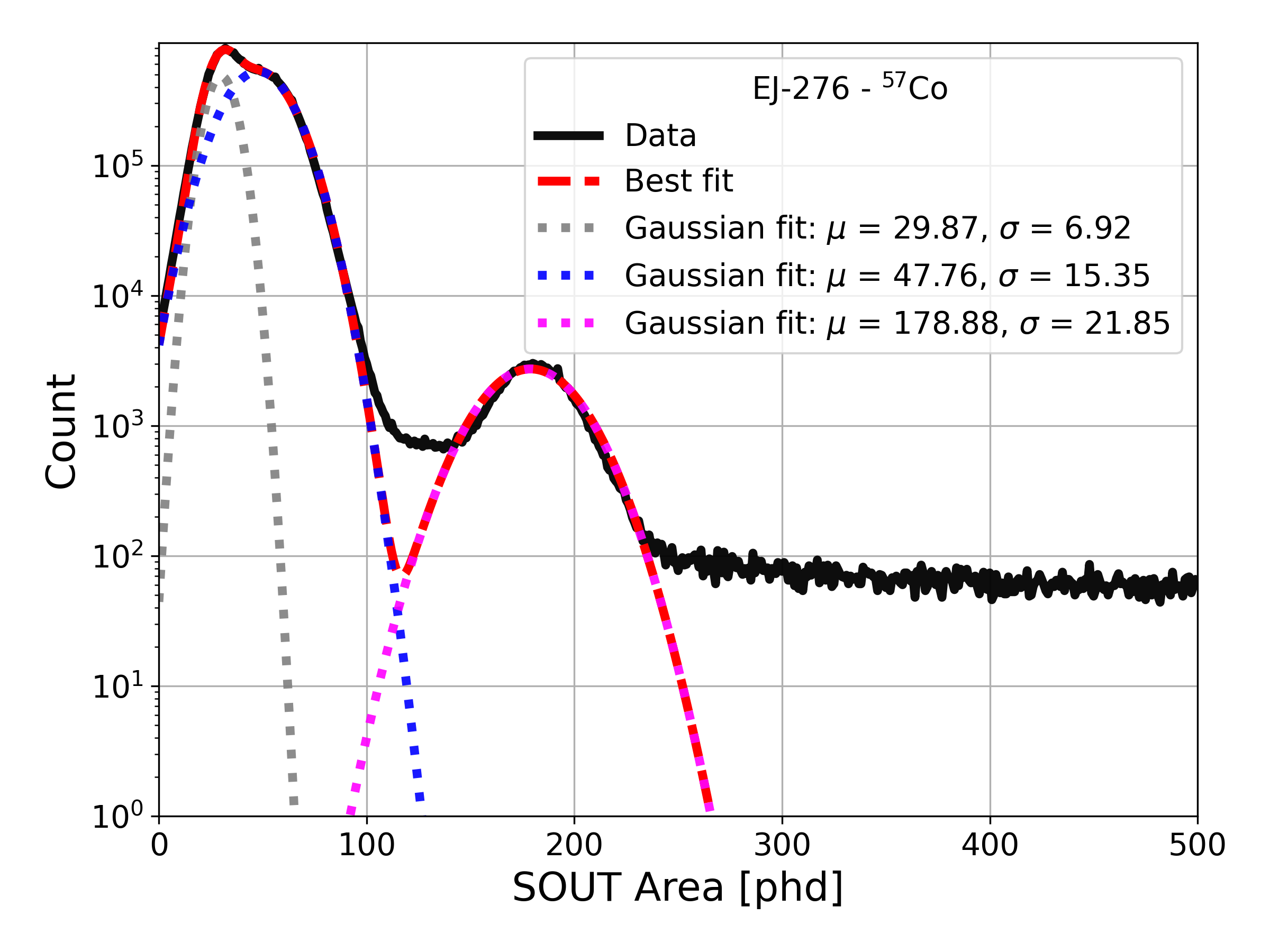}
    \includegraphics[width=0.45\textwidth]{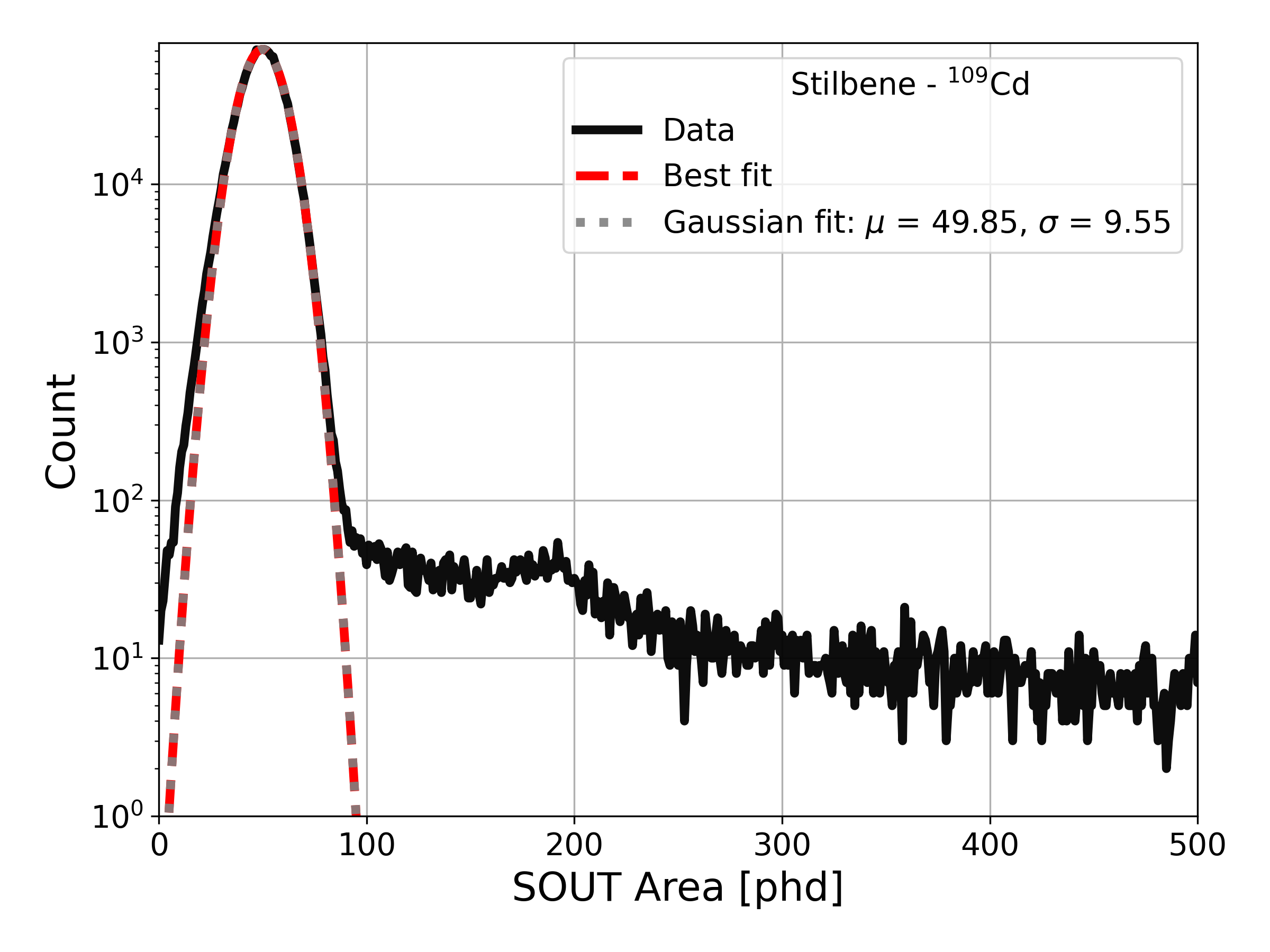}
    \includegraphics[width=0.45\textwidth]{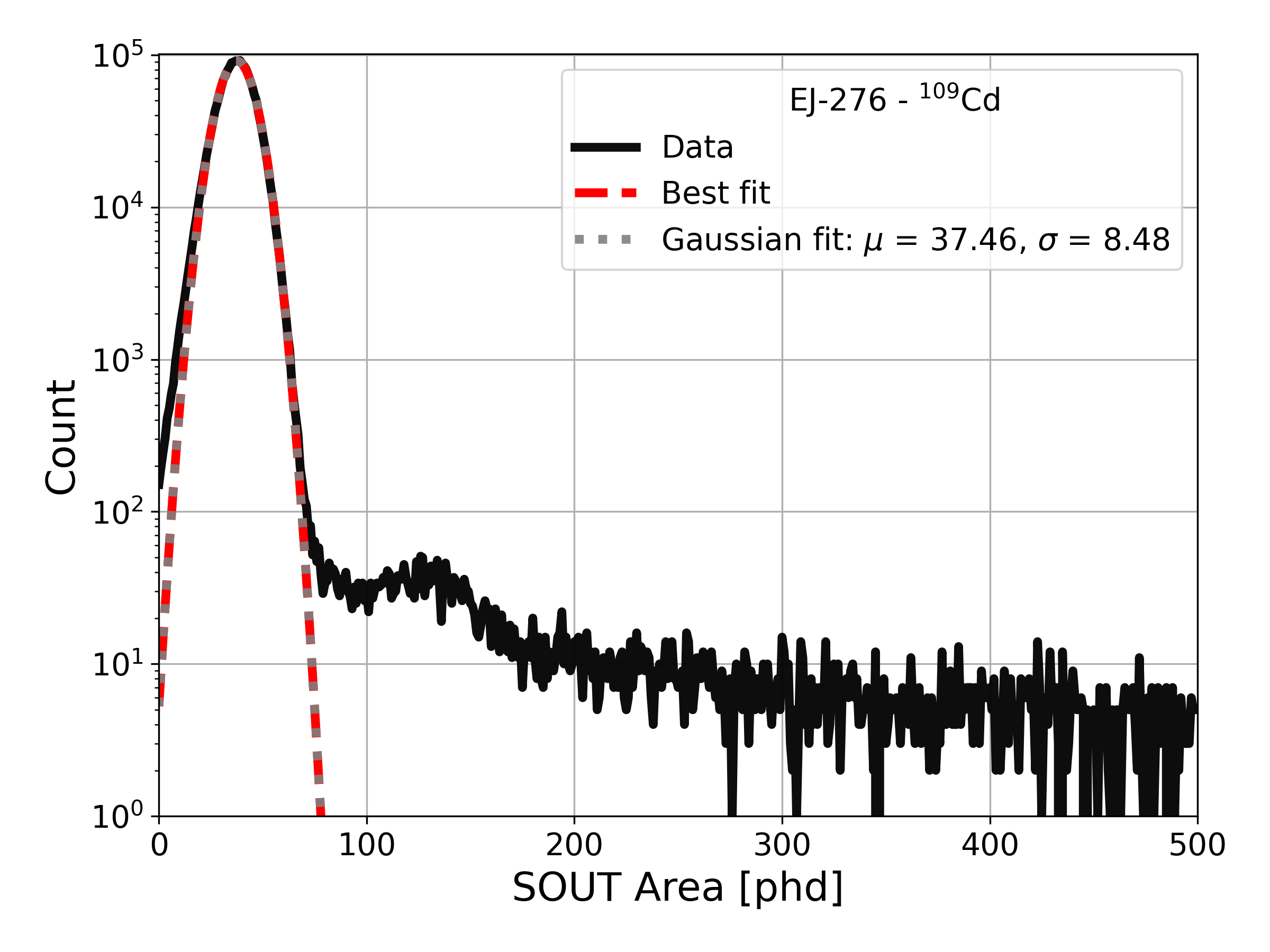}
    \includegraphics[width=0.45\textwidth]{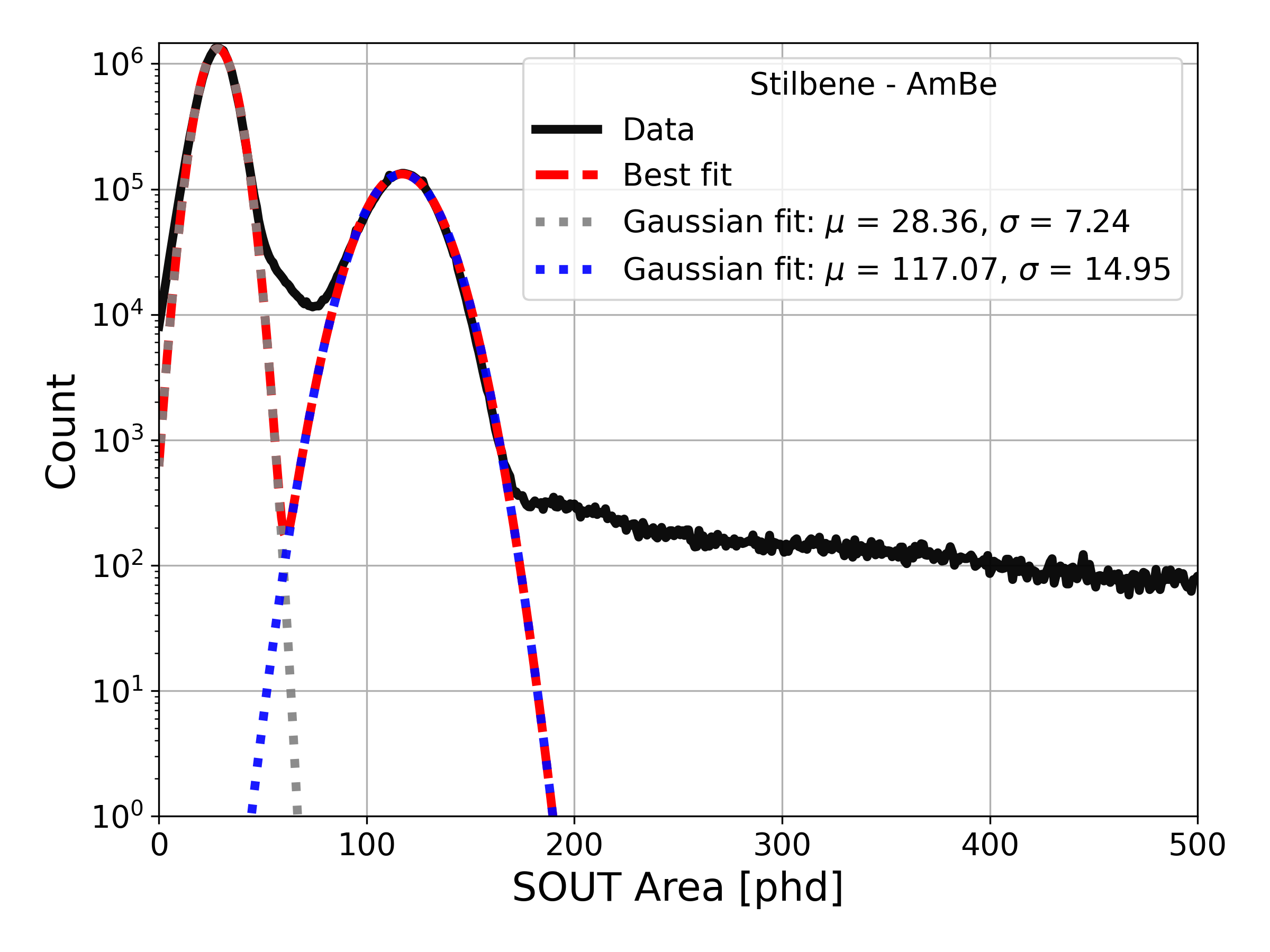}
    \includegraphics[width=0.45\textwidth]{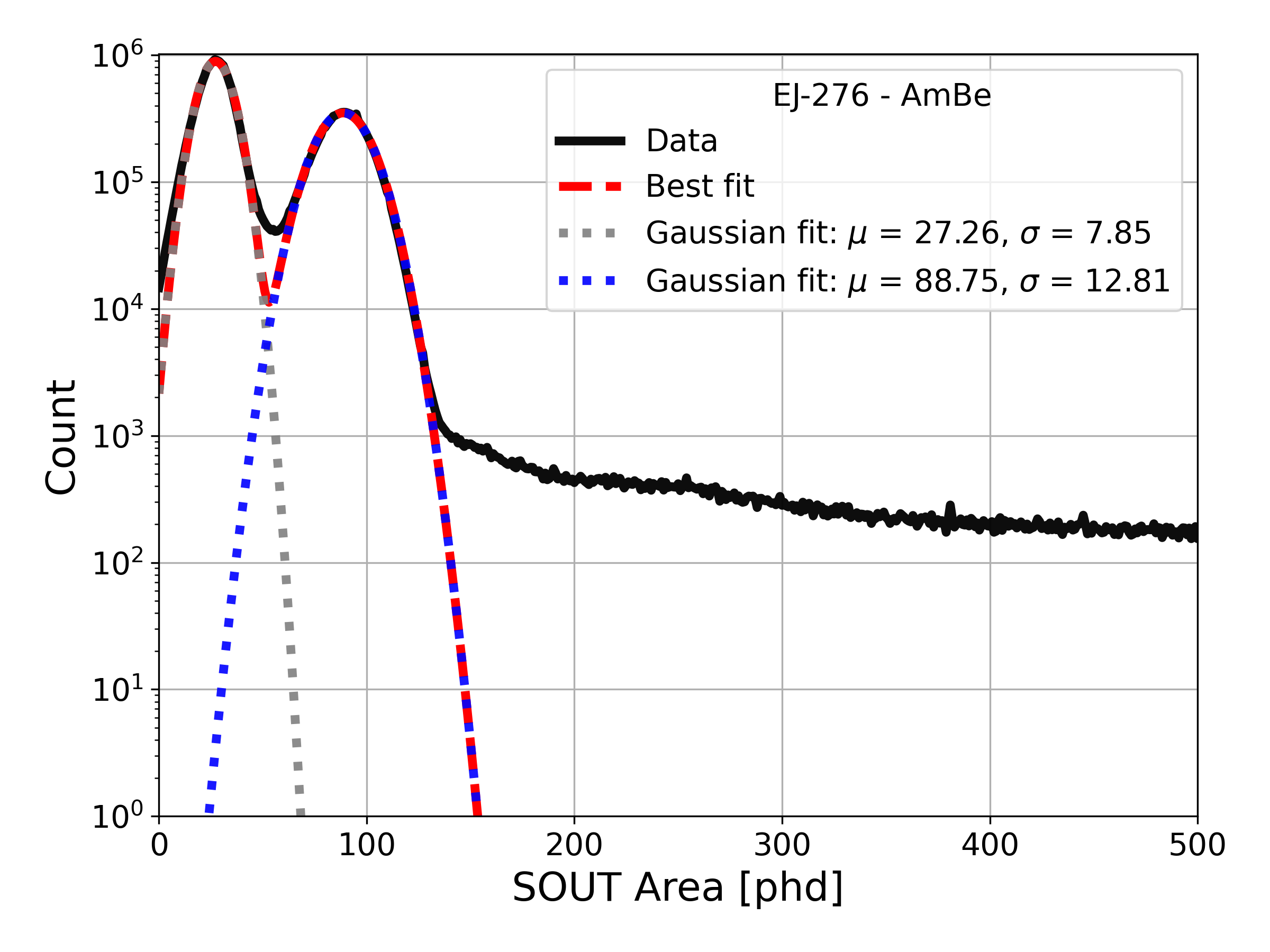}

\caption{Peak fitting of SOUT area corresponding to known gamma and X-ray emissions from $^{57}$Co, $^{109}$Cd and AmBe. The left panels show the data collected using stilbene, and the right shows the data collected using EJ-276. The mean and sigma of the individual peak fits are given in the legend.}
    \label{fig:peak_fits}
\end{figure*}

\begin{figure} [h!]
\begin{center}
\includegraphics[width=0.9\textwidth]{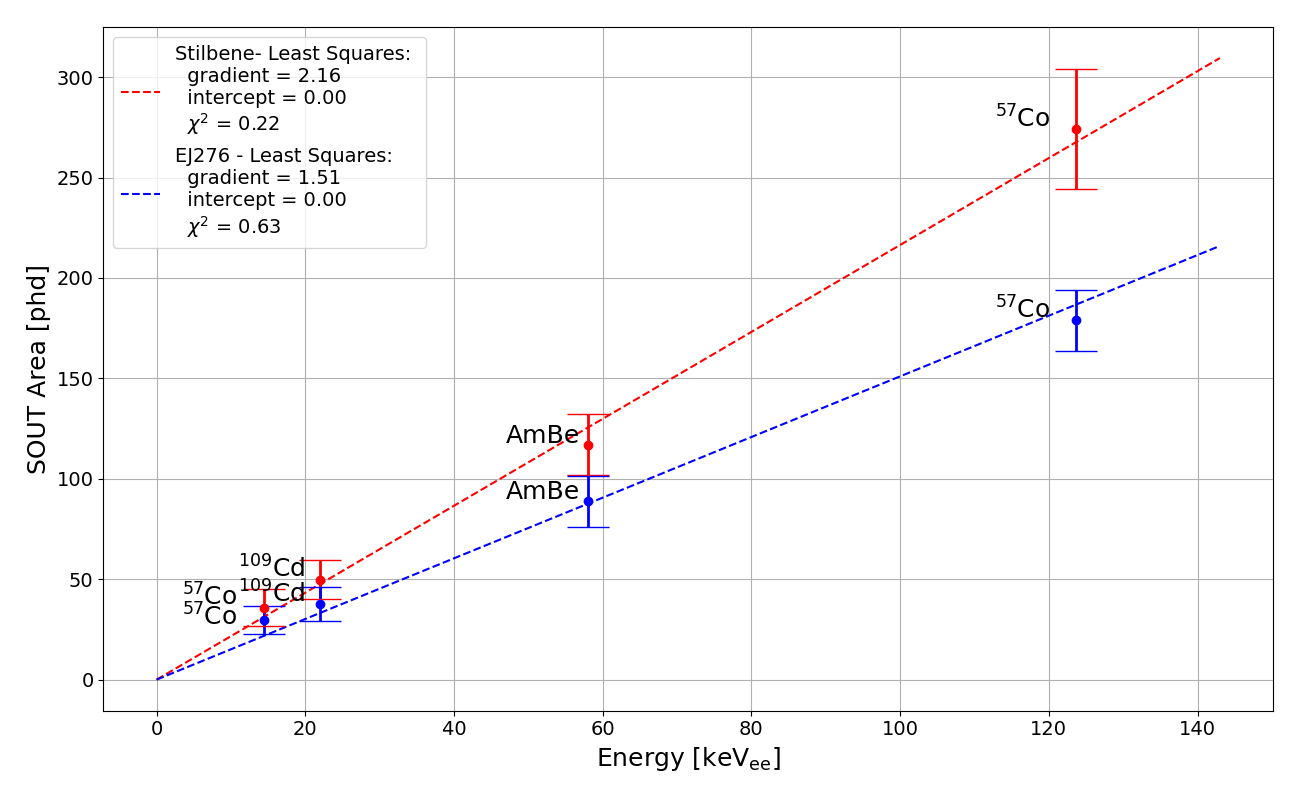}
\end{center}
\caption[example] 
{\label{fig:energy_map}Mapping of SOUT area in photons detected to energy in keV$_{\rm{ee}}$ for stilbene and EJ-276.}
\end{figure}

Two gamma-only sources,$^{57}$Co and $^{109}$Cs, as well as the 59.5 keV gamma from AmBe, a mixed neutron and gamma source, were used to determine the mapping from photons detected, phd, to deposited energy in keV electron equivalent, keV$_{\rm{ee}}$. Known peaks in the decay spectrum of each source were fit with a Gaussian, and the mean was taken as the corresponding number of phd at that energy, the results of which are given for each scintillator-source paring in Figure \ref{fig:peak_fits}.
Figure \ref{fig:energy_map} shows the least-squares linear fit performed to extract the conversion for both scintillators. For stilbene, the conversion was 2.16 phd/keV$_{\rm{ee}}$, and for EJ-276, the conversion was 1.51 phd/keV$_{\rm{ee}}$. The PSD capabilities of both scintillators were assessed using a data set obtained using the AmBe source. The entire 2 $\rm{\mu}$s acquisition window was used when determining the area of the waveforms to allow for a thorough investigation. This, however, had the adverse effect of increasing the likelihood of pile-up. Therefore, selection criteria was enforced to remove any event where the SOUT area considered out to 700 ns was not within 20\% of the area at 2 $\rm{\mu}$s. Additionally, any event with two identifiable peaks was removed. 

\section{Assessing The Optimal Integration Periods For Pulse Shape Discrimination at Low Energies}
\label{sec:results}  
\begin{figure} [htb]
\begin{center}
\includegraphics[width=0.98\textwidth,trim={ 0 0.5cm 1cm 0},clip]{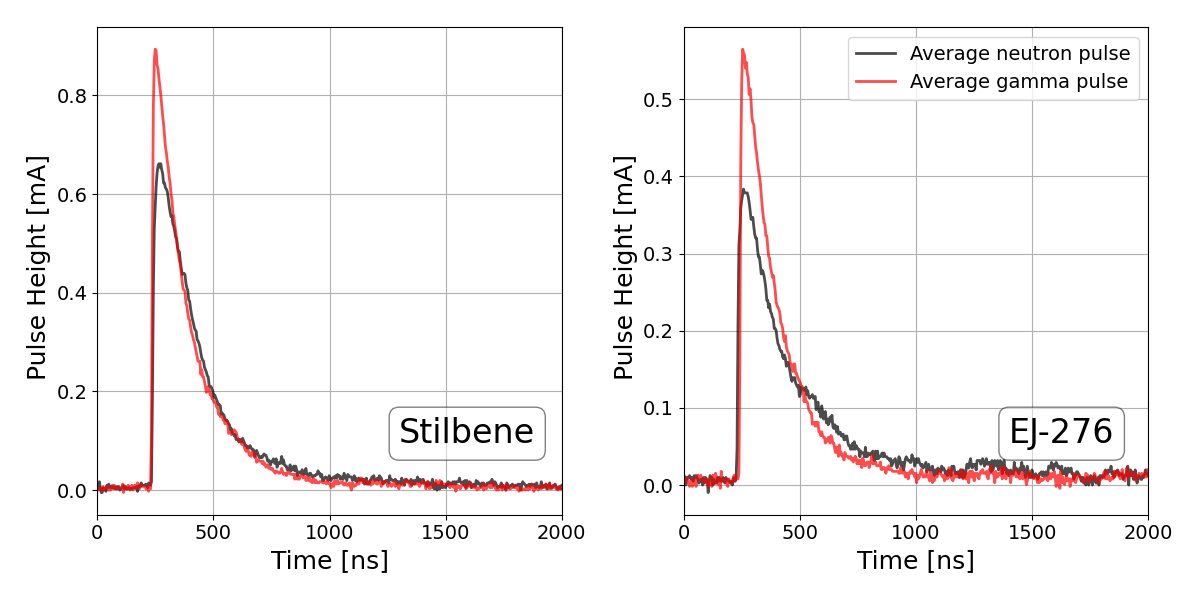}
\end{center}
\caption[example] 
{\label{fig:Waveform_100} Neutron and gamma waveforms corresponding to a 100 keV$_\mathrm{ee}$ interaction in stilbene (left) and EJ-276 (right) using an AmBe source.} \end{figure}

\begin{figure} [htb]
\begin{center}
\includegraphics[width=0.98\textwidth, trim={ 0 .5cm 4.5cm 0},clip]{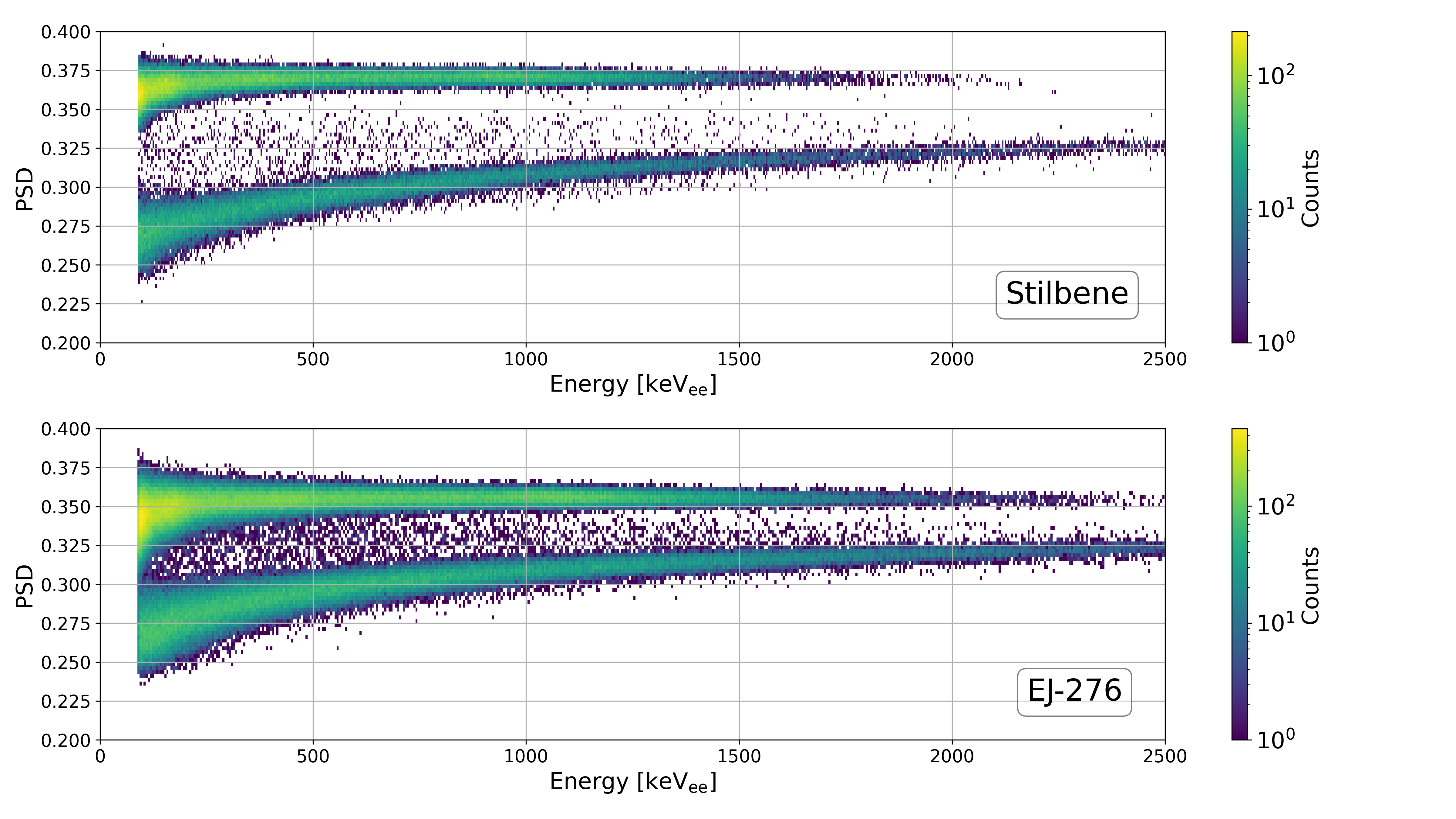}
\end{center}
\caption[example] 
{\label{fig:PSD_multiplot_SB} Distribution of the PSD metric for stilbene (top) and EJ-276 (bottom) as a function of recoil energies for a $t_1$ of 70 ns and $t_2$ of 700 ns using an AmBe. A cut on the data has been applied at 90 keV$_{\rm{ee}}$}
\end{figure}

When analyzing the data, a software-based trigger on FOUT was employed to obtain an improved definition of $t_0$. This was achieved by identifying the sample corresponding to each event's peak in the FOUT waveform using the SciPy peak finding algorithm \cite{SciPy}. The sample preceding the peak defined the $t_0$ of the acquisition; this is the sample from which the determination of the integrated outputs will start for the corresponding SOUT waveforms. 
Events with total energy less than 90 keV$_\mathrm{ee}$ were removed, giving 130k events with stilbene and 284k events with EJ-276. Typical waveforms for neutron and gamma interactions corresponding to 100 keV$_\mathrm{ee}$ in both scintillators are given in Figure \ref{fig:Waveform_100}. 
The resultant PSD values as a function of the interaction energy for both scintillators, using a $t_1$ of 70 ns and a $t_2$ of 700 ns are given in Figure \ref{fig:PSD_multiplot_SB} for events with energies greater than 90 keV$_\mathrm{ee}$.
Both scintillators display separable bands for gamma (top) and neutron (bottom). Stilbene displays better separation in PSD values between the two bands across the considered range of interaction energies.

\begin{figure} [b!]
\begin{center}
\includegraphics[width=0.98\textwidth]{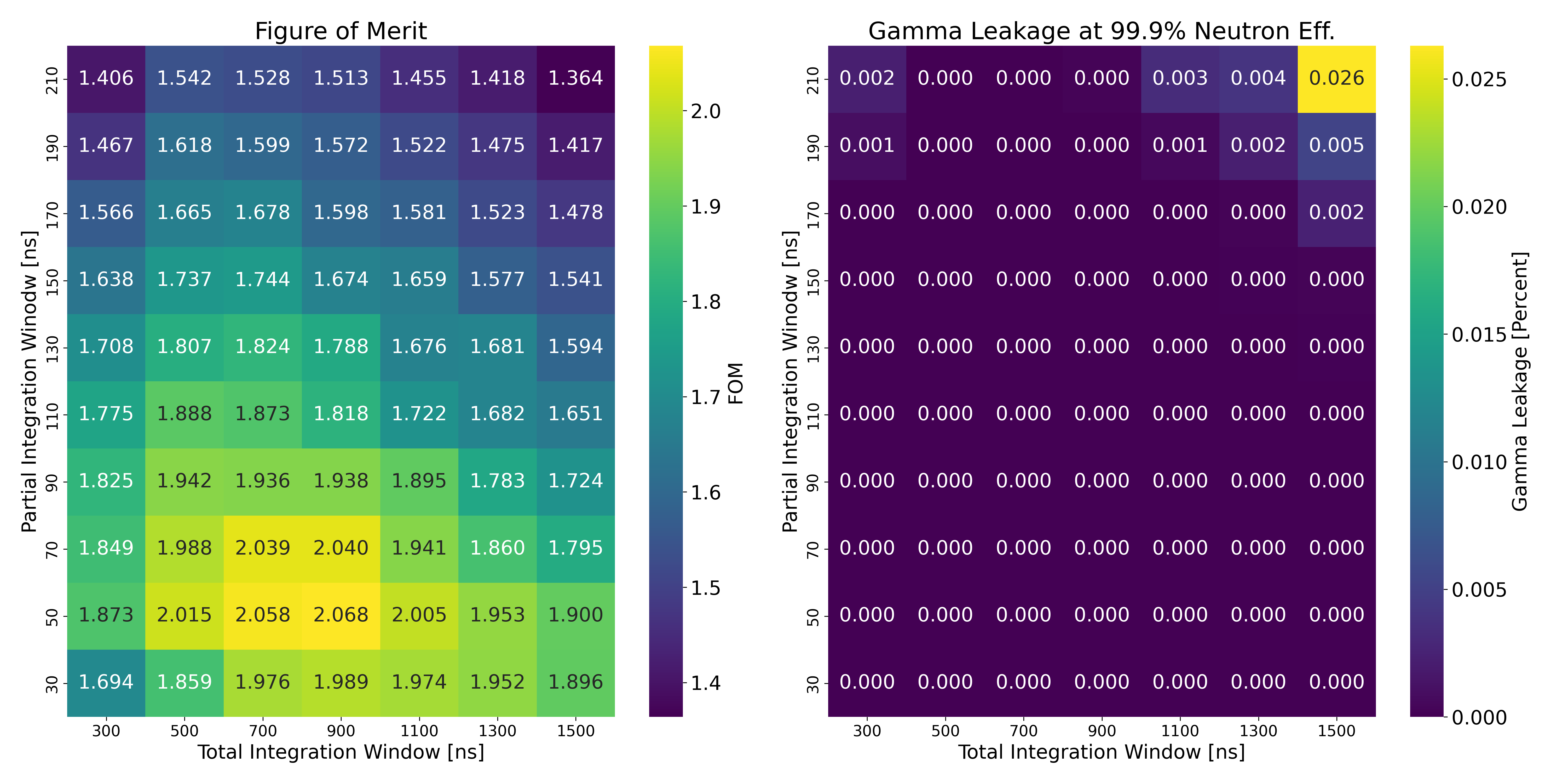}
\end{center}
\caption[example] 
{ \label{fig:FOM_GL_SB} The resultant FOM (left) and gamma leakage (right) for various partial and total integration periods for neutron and gamma interaction energies between 90 and 110 keV$_{\rm{ee}}$ in a cube of stilbene coupled to a 60035 Onsemi J-series SiPM.}
\end{figure}

\begin{figure} [t!]
\begin{center}
\includegraphics[width=0.98\textwidth]{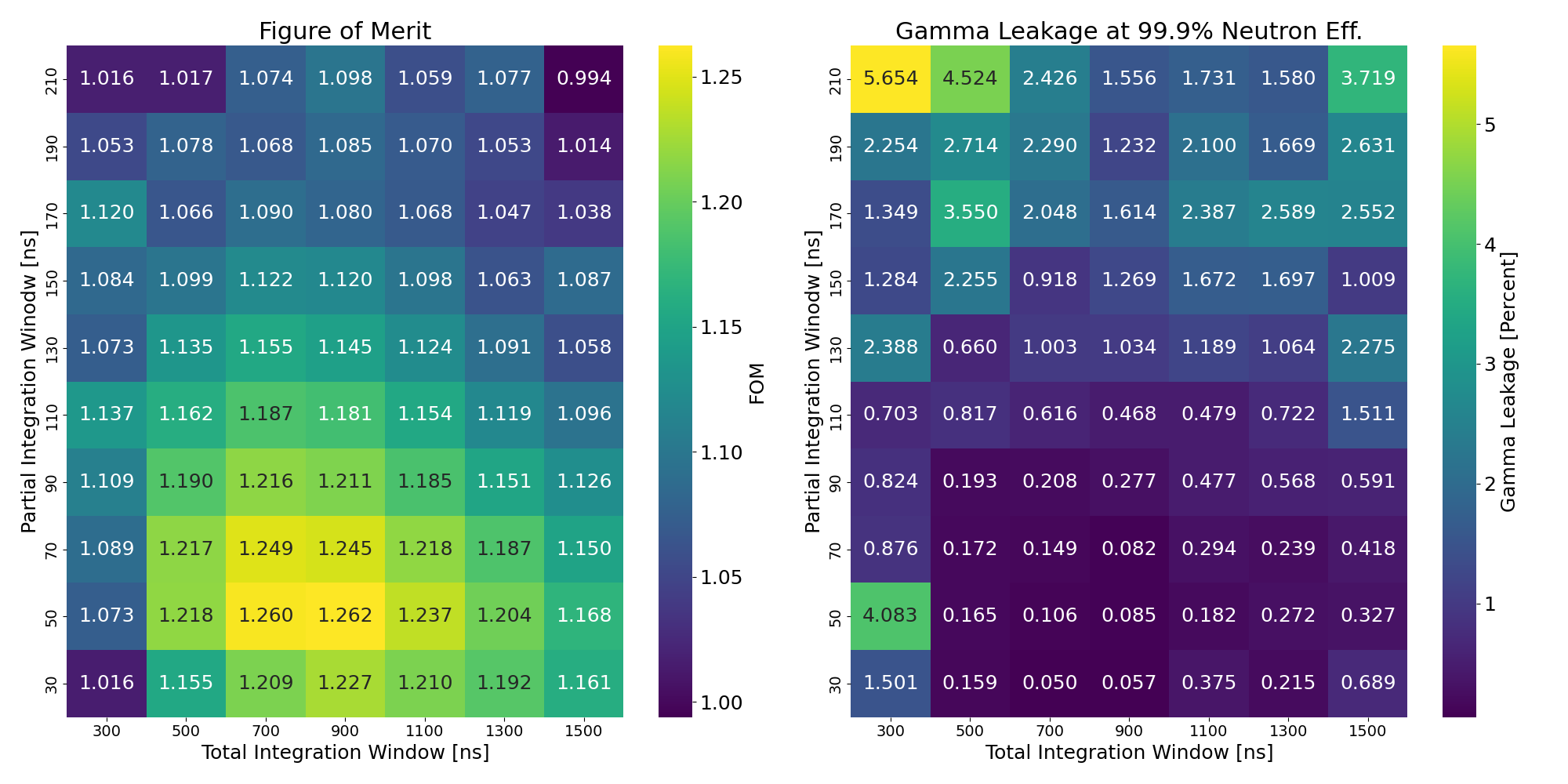}
\end{center}
\caption[example] 
{ \label{fig:FOM_GL_EJ}  The resultant FOM (left) and gamma leakage (right) for various partial and total integration periods for neutron and gamma interaction energies between 90 and 110 keV$_{\rm{ee}}$ in a cube of EJ-276 coupled to a 60035 Onsemi J-series SiPM.}
\end{figure}

The FOM and gamma leakage for both scintillators were determined for interactions with energies of 90 to 110 keV$_\mathrm{ee}$ range for combinations of partial integration periods between 30 and 210 ns and total integration periods between 300 and 1500 ns. Figure \ref{fig:FOM_GL_SB} shows the FOM and gamma leakage for different combinations of partial and total integration periods. The optimal combination for stilbene in both metrics was for a partial integration time of 50 ns and a total integration time of 900 ns, having a FOM of 2.068 and no discernible gamma leakage. For EJ-276, Figure \ref{fig:FOM_GL_EJ}, the optimal combination was for a partial integration time of 50 ns and a total integration time of 900 ns, giving a FOM of 1.262 with gamma leakage of 0.085\%, falling short of the required 1.27 threshold for separable distributions. The FOM values for stilbene indicate adequate PSD capabilities across all considered values of partial and total integration periods at energies of 100 $\pm$ 10 keV$_\mathrm{ee}$.
On the other hand, EJ-276 shows no combination meeting the required value of 1.27. However, a finer scan of partial integration from 30 to 70 ns and total integration from 500 to 900 ns might identify a more viable combination for PSD at these energies. This suggests that a device for the readout of SiPMs coupled to stilbene with a 50 ns partial and 900 ns total integration can reliably employ an on-board analog Q-ratio method to perform PSD between fast neutrons and gammas at energies as low as 100 keV. The greater separation between the two PSD distributions, and hence larger FOM values, in stilbene at these energies, is attributed partly to its higher light yield per MeV than EJ-276. It should be noted that the gamma and neutron distributions for EJ-276 do become well separated at higher energies.


\subsection{Pulse Shape Discrimination in Subtraction Space}
\begin{figure} [h!]
\begin{center}
\includegraphics[width=0.98\textwidth, trim={ 0 .5cm 4.5cm 0},clip]{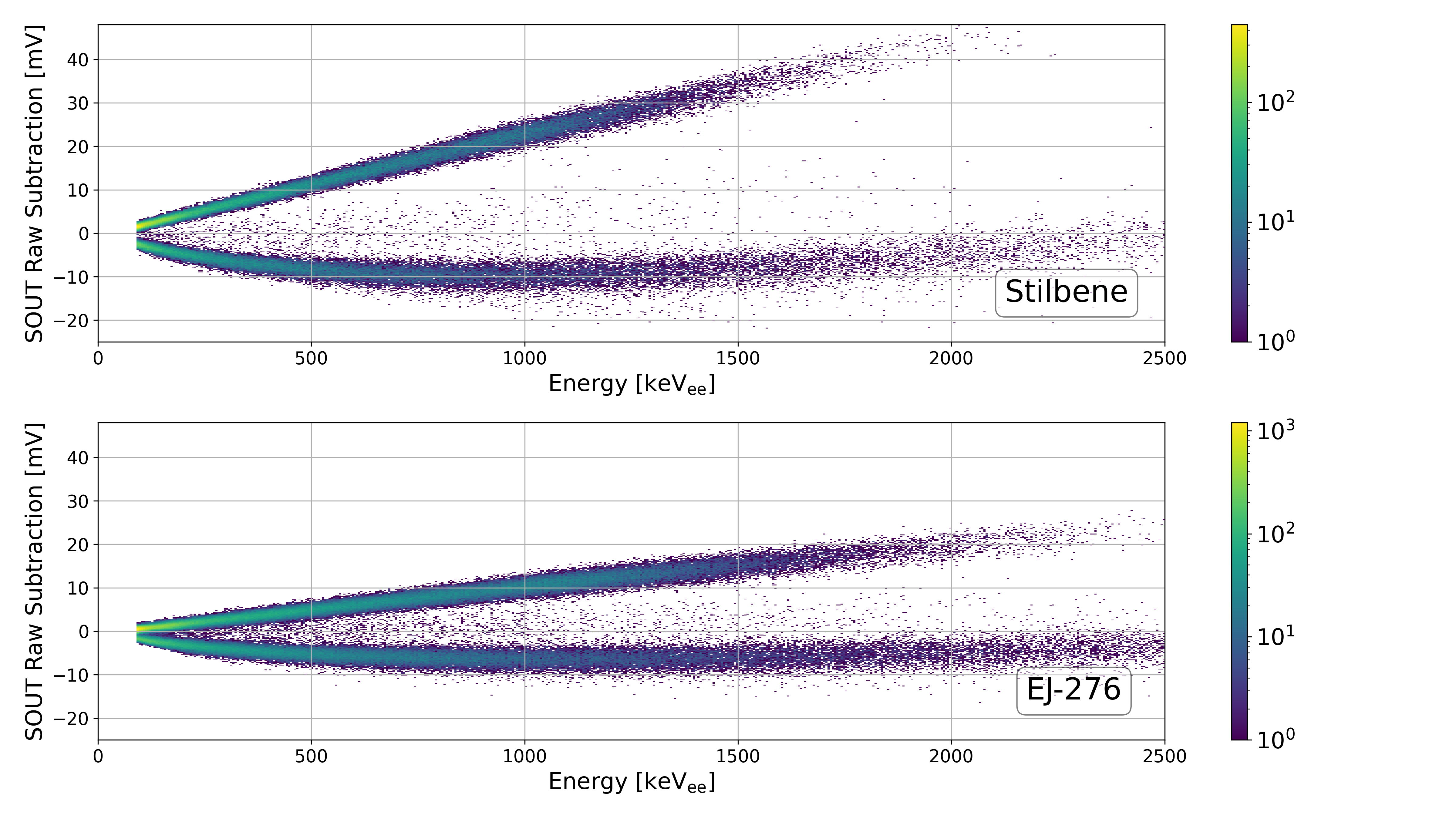}
\end{center}
\caption[example]{
\label{fig:SP_multiplot} Subtraction space for stilbene (top) and EJ-276 (bottom) as a for a $t_1$ of 70 ns and $t_2$ of 700 ns using an AmBe as a function of Energy. A cut on the data has been applied at 90 keV$_{\rm{ee}}$. Subtraction space scaling using a PSD value of 0.328 for stilbene and 0.313 for EJ-276.}
\end{figure} 

While a Q-ratio PSD approach works well in offline data processing of digitized outputs, implementing a division circuit in an ASIC is unnecessarily burdensome. Instead, a circuit involving a subtraction between partial and total integration, wherein the value of the latter is scaled, results in an equivalent circuit. A threshold in division space can be mapped to a null difference in subtraction space using this scale factor. Figure \ref{fig:SP_multiplot} illustrates this concept in the context of Figure \ref{fig:PSD_multiplot_SB}. 

\section{Conclusion}
\label{sec:conclusion}  
When applying a Q-ratio method, the optimal partial and total integration periods for interaction energies between 90 and 100 keV$_{\rm{ee}}$ in stilbene were found to be 50 ns and 900 ns, respectively, as shown by the high FOM of 2.068 and negligible gamma leakage. For interactions in this energy range in EJ-276, the FOM never met the criteria of being well separated. The combination offering the highest FOM and lowest gamma leakage was for a 50 ns partial integration and a 900 ns total integration. These values will be those used to set the bounds on the range of the integration periods required to perform onboard PSD using the ASIC.

\acknowledgments
We are thankful to Greg
Warner and Vernon Cottles of Advanced Research Corporation in Minnesota
for many discussions, and for providing the dark box that made these
measurements possible.
This work was supported by the DOE/NNSA under Award Number DE-NA0000979 through the Nuclear Science and Security Consortium. 

\bibliography{main} 
\bibliographystyle{JHEP} 

\end{document}